\documentclass{article}

\usepackage{PRIMEarxiv}

\usepackage[utf8]{inputenc} 
\usepackage[T1]{fontenc}    
\usepackage{hyperref}       
\usepackage{url}            
\usepackage{booktabs}       
\usepackage{amsfonts}       
\usepackage{nicefrac}       
\usepackage{microtype}      
\usepackage{lipsum}
\usepackage{fancyhdr}       
\usepackage{graphicx}       
\graphicspath{{media/}}     
\usepackage{amsmath,amssymb,amsfonts,amsthm}
\usepackage{algorithmic}
\usepackage{textcomp}
\usepackage{xcolor}
\usepackage{blindtext}
\usepackage{subfig}
\usepackage{cuted}
\usepackage{physics}
\usepackage{xcolor}
\usepackage[long]{optidef}
\usepackage[detect-none]{siunitx}
\usepackage{multirow}
\usepackage{array}

\usepackage{makecell}
\title{IoT Localization and Optimized Topology Extraction Using Eigenvector Synchronization 
\thanks{\textit{This material is based upon work supported by Science Foundation Ireland (SFI) and is co-funded under the European Regional Development Fund under Grant Numbers 13/RC/2077 and 13/RC/2077-P2. \\
\copyright 2023 IEEE. Personal use of this material is permitted. Permission from IEEE must be obtained for all other uses, in any current or future media, including reprinting/republishing this material for advertising or promotional purposes, creating new collective works, for resale or redistribution to servers or lists, or reuse of any copyrighted component of this work in other works.}} 
}

\author{\Large{\emph{Indrakshi Dey and Nicola Marchetti}}}

\begin{document}
\maketitle

\begin{abstract}
Internet-of-Things (IoT) devices are low size, weight and power (SWaP), low complexity and include sensors, meters, wearables and trackers. Transmitting information with high signal power is exacting on device battery life, therefore an efficient link and network configuration is absolutely crucial to avoid signal power enhancement in interference-rich environment and resorting to battery-life extending strategies. Efficient network configuration can also ensure fulfilment of network performance metrics like throughput, coding rate and spectral efficiency. We formulate a novel approach of first localizing the IoT nodes and then extracting the network topology for information exchange between the nodes (devices, gateway and sinks), such that overall network throughput is maximized. The nodes are localized using noisy measurements of a subset of Euclidean distances between two nodes. Realizable subsets of neighboring devices agree with their own position within the entire network graph through eigenvector synchronization. Using communication global graph-model-based technique, network topology is constructed in terms of transmit power allocation with the aim of maximizing spatial usage and overall network throughput. This topology extraction problem is solved using the concept of linear programming.
\end{abstract}


\section{Introduction}

Internet-of-Things (IoT) is proliferating through every possible corners of our lives with applications and services that improves and uplifts the quality of lives across different aspects \cite{1}. Localization is an essential process in the IoT environment for tracking and monitoring the targets with the help of sensor nodes. Localization is also of paramount importance for optimizing the spatial usage and network coverage with minimum number of devices \cite{2}.

IoT networks are a collection of low size, weight and power (SWaP) devices distributed over a geographical area and under different radio conditions that cooperate to monitor and manage various physical or environmental condition. Such devices are capable of limited computing power, memory, communication capabilities. Therefore, many IoT nodes cannot afford Global Navigation Satellite Systems (GNSS) \cite{3,4} receivers, which is the most common method of wireless device localization, due to their high power and cost requirements. Localization in IoT networks is also challenging owing a) complexity of the surrounding radio conditions, b) large amount of sensor errors resulting from low-cost sensors and c) dynamic nature of IoT nodes for certain applications \cite{5,6,7,8,9,10,11,12}.

Different database-matching (DB-M)-based techniques have been also introduced for localization \cite{13,14,15}. Machine Learning (ML)-based DB-M techniques like Artificial Neural Networks (ANN) \cite{16,17}, random forests \cite{18,19}, Deep Reinforcement Learning (DRL) \cite{20,21}, have also been introduced for IoT node localization. However, any DB-M technique will need a strong database which may be difficult to get hold off when deploying a new network. Moreover ML-based techniques are always computationally heavy and are not feasible for IoT nodes with limited battery life.

In this paper, we formally argue that IoT node localization and network topology extraction can be optimally achieved by investigating the relationship between the spatio-temporal structure of a network and flow of information over the network links. We embed graph over the network structure, break the large network graph into many small overlapping sub-graphs or patches and stitch them together consistently in a global coordinate system. A communication graph-model-based topology of the IoT network is constructed in this way with the aim of maximizing the spatial usage. In this respect, we investigate the question whether there exists certain power assignment that enables topology control for maximizing the spatial usage. 

To enable effective topology extraction, we formulate a signal-to-noise ratio (SNR)-based optimization problem that can be solved to obtain the minimal possible power that each IoT node can fire with, to reach its farthest neighbor (as defined in the communication graph). By iteratively invoking our formulated topology extraction, the power assignment converges within 7 iterations, on average, to an operational point that maximizes network throughput. Our proposed `\emph{Topology Extraction for Maximizing Network Throughput (\textbf{MaxNTtop})}' improves achievable network throughput by around 45\% over the traditional Brute-force search-based algorithm.

The remainder of the paper is organized as follows. We first introduce in Section II the graph embedding algorithm for mapping the spatio-temporal relationship and flow of information within an IoT network. Following that, we investigate in Section III the issue of whether or not a feasible power assignment
exists that enables the communication graph capable of satisfying overall network lower SNR threshold. Towards this end, we devise a new topology extraction algorithm, called \textbf{MaxNTtop} that iteratively invokes topology formulation until the power assignment converges to an operational point. We present in Section IV a simulation study and show the superiority of the topology induced by \textbf{MaxNTtop} in terms of maximizing the network throughput and spatial usage. Finally, we conclude the paper in Section V.

\section{Graph Embedding}

\begin{figure}[t]
\begin{center}
 \includegraphics[width=0.5\linewidth]{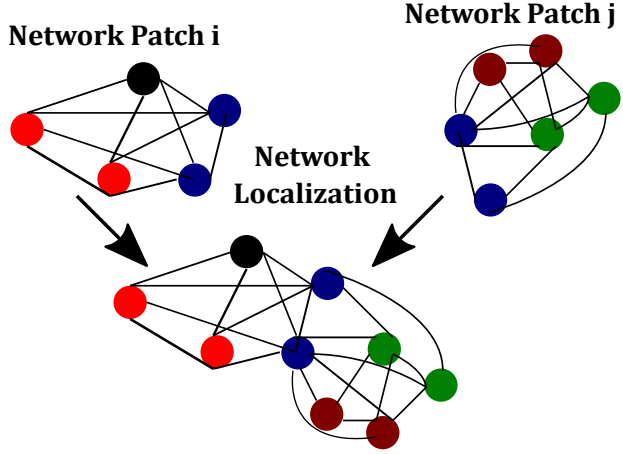}
\end{center}
\vspace*{-3mm}
\caption{A Conceptual representation of our graph embedding algorithm for aligning patches or sub-graphs within an IoT network with the over-arching goal of maximizing network coverage.}
\label{FIG1}
\vspace*{-5mm}
\end{figure}

An IoT network is modelled as an oriented graph $G = (V,E)$ where $|V| = n$ and $V$ is the set of IoT nodes, while $|E| = m$ and $E$ is the set of links (edges) which is associated with a distance measurement matching the Euclidean distance between two neighboring IoT nodes. In other words, the distance between nodes $i$ and $j$, $d_{ij} = d_{ji}$ and the link (edge) between $(i,j) \in E$. For any such graph realization problem, the main goal is to find graph embeddings $p_1, p_2, \dotso, p_n \in \mathbb{R}^{\mathsf{d}}$, in $\mathsf{d}$-dimensional space and for any distance $d_{ij}$ and link between $(i,j) \in E$, where the embeddings will satisfy, $||p_i - p_j || = d_{ij}$. For realistic embeddings, we define a noisy representation of the distance, $d_{ij} = ||p_i - p_j || + \varepsilon_{ij}$ where $\varepsilon_{ij}$ is the added noise, and we have to find the optimal graph embedding considering $\varepsilon_{ij}$. The key idea behind our contribution is to divide $G$ into smaller sub-graphs which can be sewed together back in the global graph representing the entire IoT network, in a way that the number of fold-overs is minimized and both local sub-graphs and global graph are globally rigid\footnote{A globally rigid graph refers to a graph in planar Euclidean space which is structurally stable, i.e. the edges can be replaced by solid rods. Evaluating rigidity of a graph is a separate detailed mathematical study, which is beyond the scope of this paper. The verifiability and rigidity of the graph embedding outlined in this paper will be studied elaborately in our future work.}. When we are sewing the sub-graphs or patches, we need to find the reflection, rotation and translation for maximally aligning two sub-graphs in a consistent manner. This process is referred to as \emph{synchronization} \cite{22} in traditional graph theory. A conceptual overview of our graph embedding is presented in Fig.~\ref{FIG1}.

\subsection{Graph Embedding Algorithm}

For every node $j$, $V(i) = \{j : (i,j) \in E\} \cup \{i\}$ represent the set of neighbors and $G(i) = (V(i),E(i))$ represents the sub-graph consisting of 1-hop neighbors of $i$ (i.e. the set of IoT nodes which can be reached from node $i$ in just one hop; we refer to those nodes as the 1-hop neighbors). We will embed $G(i)$ in $\mathbb{R}^2$ if $G(i)$ is globally rigid and we will break $G(i)$ into more sub-graphs if $G(i)$ is not globally rigid. Therefore, to summarize our algorithm, we divide the network graph $G$ into $N$ globally rigid sub-graphs or patches denoted by $P_1, \dotso, P_N$ and then embed each of those patches $P_i$ separately using a graph embedding method of our choice.

\subsubsection{Patch Reflection}

Our aim is to estimate the set of rigid motions, $\sigma_1, \dotso, \sigma_N$, where the $i$th motion $\sigma_i$ moves the patch $P_i$ to its correct position with respect to the global coordinate system and $\sigma_i \in \text{Euc}(2)$ where $\text{Euc}(2)$ is the Euclidean group of rigid motions on $\mathbb{R}^2$. Let us consider that we want to align the patches $P_i$ and $P_j$ where $P_i$ and $P_j$ have more than two IoT nodes in common. Let us also consider that the relative reflection can be estimated as, $z_{ij} \in \{-1,+1\}$, of the patches $P_i$ and $P_j$. Using $z_{ij}$ , we build a $N \times N$ sparse symmetric matrix $Z$ such that,
\begin{align}
    Z = (z_{ij}) =
  \begin{cases}
    1   & \quad \text{align}~P_i~\text{\&}~P_j~\text{without any reflection}\\
    -1  & \quad \text{align}~P_i~\text{\&}~P_j~\text{with one reflection}\\
    0  & \quad P_i~\text{\&}~P_j~\text{cannot be aligned}
  \end{cases}
\end{align}
The next step then is to compute the top eigenvector\footnote{The top eigenvector is defined as the direction onto which a projection of the principal components has the largest variance.} $v^{\zeta}_1$ of $\zeta = \Delta^{-1}Z$ where $\Delta$ is a diagonal matrix with $\Delta_{ii} = \text{deg}(i)$ satisfying $\zeta v^{\zeta}_1 = \lambda^{\zeta}_1 v^{\zeta}_1$. Now using the eigenvector quantification and the reflections $z_{ij}$, we can estimate the global reflection of patch $P_i$, 
\begin{align}
\hat{z}_i = \text{sign}(v^{\zeta}_1(i)) = v^{\zeta}_1(i)/|v^{\zeta}_1(i)|
\end{align}
and for the cases where $\hat{z}_i = - 1$, $P_i$ will be replaced by its mirrored image.

\subsubsection{Patch Rotation}

Similar to the reflections, we will estimate the relative rotations, $\theta_{ij}$, of the patches $P_i$ and $P_j$, such that $\theta_{ij} \in [0,2\pi)$ and $r_{ij} = e^{\imath \theta_{ij}}$. Using $r_{ij}$, we construct another $N \times N$ Hermitian matrix, $R = (r_{ij})$, such that,
\begin{align}
 r_{ij} =
  \begin{cases}
    e^{\imath \theta_{ij}} & \quad P_i~\text{\&}~P_j~\text{can be aligned}\\
    0  & \quad P_i~\text{\&}~P_j~\text{cannot be aligned}
  \end{cases}
\end{align}
where since $R$ is a Hermitian matrix, $\theta_{ij} = - \theta_{ji} \mod 2\pi$ and it follows that $R_{ij} = \bar{R}_{ji}$ for any complex number $\omega = a + \imath b$ and its complex conjugate, $\bar{\omega} = a - \imath b$. We also compute the top eigenvector $v^{\mathcal{R}}_1$ of $\mathcal{R} = \Delta^{-1}R$ such that $v^{\mathcal{R}}_1$ satisfies $\mathcal{R} v^{\mathcal{R}}_1 = \lambda^{\mathcal{R}}_1 v^{\mathcal{R}}_1$. This can be used to estimate the global rotation angle $\hat{\theta}_i$ of patch $P_i$ using,
\begin{align}
e^{\imath \hat{\theta}_{i}} = v^{\mathcal{R}}_1(i)/|v^{\mathcal{R}}_1(i)|
\end{align}
and with the estimated angle $e^{\imath \hat{\theta}_{i}}$ we can rotate the patch $P_i$ for embedding. 

\subsubsection{Patch Translation}

The final step is to compute coordinates (2-dimensional) of each patch when sewed back into the global graph. Let us consider the $k$th patch $P_k$ with associated graph $G_k = (V_k, E_k)$ where $G_k$ needs to be embedded into $G = (V,E)$. For the $i$th node in patch $P_k$, we should have,
\begin{align}\label{eq5}
p_i = p_i^{(k)} + t^{(k)}; i \in V_k; k = 1, \dotso, N
\end{align}
where $t^{(k)}$ is the associated translation. Using the over-determined system of (\ref{eq5}), we can estimated the global coordinates, $p_1, \dotso, p_n$ using least square solution to (\ref{eq5}). The set of translations $t^{(1)}, \dotso, t^{(N)}$ is disregarded. So for each edge $(i,k) \in E_k$, $p_i - p_j = p_i^{(k)} - p_j^{(k)}$. If we replace the patches with the $(x,y)$ coordinates, $x_i - x_j = x_i^{(k)} - x_j^{(k)}; y_i - y_j = y_i^{(k)} - y_j^{(k)}$, which can be solved separately. Now we can write, $\tau x = \gamma^x$, where $\tau$ is the least square matrix, $x$ is the $n \times 1$ vector of $x$-coordinates of all IoT nodes localized in a patch or sub-graph, and $\gamma^x$ is the vector with entries from the right-hand side of the last equation. Similarly, we can write $\tau y = \gamma^y$. Now by adding all the equations corresponding to the same edge $(i,j)$ from different patches,
\begin{align}\label{eq6}
&\sum_{k \in \{1, \dotso, N\}; (i,j) \in E_k} x_i - x_j \nonumber\\
&= \sum_{k \in \{1, \dotso, N\}; (i,j) \in E_k} x_i^{(k)} - x_j^{(k)},~(i,j) \in E 
\end{align}
and similarly for the $y$-coordinates we can formulate the least-square transition matrix $\tau$, i.e. $m \times n$ over-determined system of linear equations. If the estimated transition matrix is denoted by $\hat{\tau}$, the least square solutions to $\hat{\tau}x = \gamma^x$ and $\hat{\tau}y = \gamma^y$ will be given by, $\hat{p}_1, \dotso, \hat{p}_n$. The error in any estimated patch coordinates can be calculated as $||p_i - \hat{p}_i||$. 

\subsection{Patch Embedding for globally non-rigid sub-graph}

For patch embedding in our globally non-rigid IoT network sub-graphs, we resort to the three-step method outlined in \cite{23}. The first step is to estimate the missing distance, $d'_{ij}$ with $(i,j) \notin E_k$,\footnote{We are considering an annular graph, with a network where the IoT devices are distributed randomly on an annulus surrounding the IoT gateway at the center.}
\begin{align}\label{eq7}
d'_{ij} = (\underline{d_{ij}} + \overline{d_{ij}})/2 
\end{align}
for,
\begin{align}\label{eq8}
\overline{d_{ij}} &= \min_{k:(i,k),(j,k) \in E_k} d_{ik} + d_{jk}\\
\underline{d_{ij}} &= \max\bigg[\max_{k:(i,k)\in E_k}d_{ik}, \max_{k:(j,k)\in E_k}d_{jk}\bigg]
\end{align}

The second step is to compute local coordinates of all nodes in a patch using multi-dimensional scaling \cite{24} on the complete set of pairwise distances. We can express, $\mathbf{D} = -1/2 \mathbf{J}\mathbf{L}\mathbf{J}$ where $\mathbf{J} = \mathbf{I}_n - 1/n \mathbf{1}\mathbf{1}^t$, $\mathbf{1}$ is a matrix of ones, $\mathbf{I}$ is the identity matrix, $\mathbf{L}$ is the matrix of squared pairwise distances, $\mathbf{L} \in \mathbb{R}^{n \times n}$, $t$ denotes the transpose and $n \geq 1$ is an integer and the matrix of the local coordinates of all nodes in a patch can be calculated as,
\begin{align}\label{eq10}
&P_k = \mathbf{U}_k \sqrt{\Lambda_k} \nonumber\\
&\quad = \bigg[\sqrt{\lambda_1}\mathbf{u}_1, \dotso, \underbrace{\sqrt{\lambda_{r+1}}\mathbf{u}_{r+1}}_{= 0}, \dotso, \underbrace{\sqrt{\lambda_k}\mathbf{u}_k}_{= 0}\bigg] \in \mathbb{R}^{n \times k}
\end{align}
where $({\lambda_n},\mathbf{u}_n)$ are the eigen-pairs of the $\mathbf{D}$ matrix.

The final step is to refine the embedding using iterative majorization technique. The coordinates of each IoT node are updated according to,
\begin{align}\label{eq11}
&p_i \leftarrow 1/~\text{deg}_i(P_k) \nonumber\\
&\quad \sum_{j \in V_k, (i,j) \in E_k} [p_j + d_{ij}(p_i - p_j)~\text{inv}~(||p_i - p_j||)]
\end{align}
where $\text{deg}_i(P_k)$ denotes the degree of node $i$ in patch $P_k$ and,
\[ \text{inv}~(x) =
  \begin{cases}
    1/x       & \quad \text{if } x \neq 0\\
    0  & \quad \text{if } x = 0
  \end{cases}
\]

\subsection{Aligning Patches and Sewing them together}

Following the embedding of the patch $P_k$, now let us look into how we can align $P_k$ with another patch $P_l$ where $P_k$ and $P_l$ have more than 2 IoT nodes in common. Let us denote the set of intersecting nodes between $P_k$ and $P_l$ as $V_{k,l} = \{v_1, \dotso, v_s\} = V_k \cap V_l$, the coordinates of $V_{k,l}$ in $P_k$ as $p_1^{(k)}, \dotso, p_s^{(k)}$, and the coordinates of $V_{k,l}$ in $P_l$ as $p_1^{(l)}, \dotso, p_s^{(l)}$. For the $i$th node in $P_k$, $p_i^{(k)} = (x_i^{(k)}, y_i^{(k)}) = x_i^{(k)} + \imath y_i^{(k)}$ is mirrored across to the $x$-axis to the rotation $\bar{p}_i^{(k)} = (x_i^{(k)}, - y_i^{(k)}) = x_i^{(k)} - \imath y_i^{(k)}$. The aim is to find $r_{\theta} = e^{\imath \theta}$ and translation vector, $\tau = x + \imath y$, that minimize the following objective function,
\begin{align}\label{eq12}
f(\theta,\tau) = \sum_{i =1}^s |p_i^{(k)} - (r_{\theta} p_i^{(l)} + \tau)|^2
\end{align} 
Now $P_l$ may be aligned with $P_k$ in its original form or we may need to consider the mirror image of $\bar{P}_l$ for aligning. The objective function that needs minimization for mirrored $P_l$ is,
\begin{align}\label{eq13}
\tilde{f}(\theta,\tau) = \sum_{i =1}^s |p_i^{(k)} - (r_{\theta} \bar{p}_i^{(l)} + \tau)|^2
\end{align} 
Consequently, we can define the required relative reflection, $z_{ij}$, required for final aligning as,
\begin{align}\label{eq14}
   z_{kl} =
  \begin{cases}
    1   & \quad \text{if}~\min_{\theta,i} f(\theta,\tau) \leq \min_{\theta,i} \tilde{f}(\theta,\tau)\\
    -1  & \quad \text{if}~\min_{\theta,i} f(\theta,\tau) > \min_{\theta,i} \tilde{f}(\theta,\tau)\\
    0  & \quad P_k~\text{\&}~P_l~\text{cannot be aligned}
  \end{cases}
\end{align}
We can then rewrite (\ref{eq12}) and (\ref{eq13}) as $||A\xi - B||^2$ where,
\begin{align}\label{eq15}
&A^t = 
 \begin{pmatrix}
  p_1^{(l)} & \dotso & p_i^{(l)} & \dotso & p_s^{(l)} \\
  1 & \dotso & 1 & \dotso & 1
 \end{pmatrix}\nonumber\\
&B^t = (p_1^{(k)} \dotso p_i^{(k)} \dotso p_s^{(k)});~\xi = [r_{\theta},\tau]
\end{align} 
in order to solve the minimization problem in (\ref{eq13}) for the rotation and translation vector for the patches. The optimal solution is obtained by solving the problem in (\ref{eq12}) using complex least squares following the method outlined in Subsection II-C.

\section{Topology Extraction}

Let us now consider two IoT nodes $V_i$ and $V_j$ with Euclidean distance $d_{ij} = d(v_i, v_j)$ and Euclidean coordinates $v(x,y), x \in X, y \in Y$. Let every node $v_i$ fire with a transmit power $\rho_T(i)$ and all the transmit power assignments are collated in, $\{\rho_T(1), \dotso, \rho_T(n)\}$. On the receiver side, we can define the signal-to-noise ratio (SNR) of a reliably detectable signal as,
\begin{align}\label{eq16}
\text{SNR}_{ij} = h_{ij}~\rho_T(i)~d_{ij}^{-\nu}/\mathcal{N} \geq \beta 
\end{align} 
where $h_{ij}$ is the channel gain (link gain) between the $i$th and $j$th IoT nodes, $\nu$ is the pathloss exponent, $\mathcal{N}$ is the noise power and $\beta$ is the lower SNR threshold or the minimum detectable SNR on the receive side.

With the constraint (\ref{eq16}) in place, we can now formulate the problem of power assignment for maximizing spatial usage or network coverage such that the network throughput is maximized. Specifically, we model the problem as a system of linear programming with respect to transmit power $\rho_T(i), i = 1, \dotso, n$:
\begin{align}\label{eq17}
\text{minimize}~~\sum_{i = 1}^n &\rho_T (i) \nonumber\\
\text{subject to,}~~&\rho_T(i) \leq \rho_{T_{\text{max}}};~\text{SNR}_{ij} \geq \beta \nonumber\\
&\rho_T(i) \geq d_{ij}^{\nu}~\rho_{R_{\text{min}}}~~ \forall e_{ij} \in G
\end{align} 
where $\rho_{T_{\text{max}}}$ is the maximum power with which any IoT device can fire a signal (depending on the particular device battery life) and $\rho_{R_{\text{min}}}$ is the minimum received power which is detectable and decodable by the receiver (gateway or any other IoT network device).

Next we will be using the solution to the power allocation problem formulated in (\ref{eq17}) and improve network throughput. First we compute power assignment to maximize spatial usage and then extract the corresponding network topology. The algorithm continues until it converges to a point where network throughput is maximized. We summarize the algorithm as below.\\

\noindent\textbf{Topology Extraction for Maximizing Network Throughput (\textbf{MaxNTtop})}\label{alg:cap}\\

\textbf{Input} : A set of IoT nodes $V$ and their coordinates $\{X,Y\}$\\
Let us consider an iteration step $\varepsilon$ of a small value\\
\textbf{Step 1} : Power assignment $\mathsf{P}_T \to \{\rho_T(1), \dotso, \rho_T(n)\}$\\ 
For all node pairs $i,j$ such that $d_{ij} \leq$ transmission range, compute SNR by (\ref{eq16})\\
\textbf{Step 2} : Sort edges in the decreasing order of SNR\\
Let $\hat{e}_1, \hat{e}_2, \dotso$ be the resulting sequence of edges, initialize $m$ patches\\
If patch ($i$) $\neq$ patch ($j$) then sew them using graph embedding algorithm outlined in Section II, $E = E \cup \{\hat{e}_{i,j}\}$, where $\hat{e}_{i,j}$ is the newly formed edge after patch alignment\\
\textbf{Extract Topology} $\Gamma(V,E)$ \\
\textbf{Step 3} : Calculate SNR$(\Gamma, \mathsf{P}_T)$ and initialize $\Delta = 1$\\
For all; $\Delta > \varepsilon$; SNR$_{\text{max}}$ = SNR$(\Gamma, \rho_{T_{\text{max}}})$; $\Delta = ||\text{SNR}_{\text{max}} - \text{SNR}(\Gamma, \mathsf{P}_T)||$\\
\textbf{Output} : Finalize power assignment $\mathsf{P}_T$
It is worth-mentioning here that the number of iterations required for \textbf{MaxNTtop} to converge is independent of the network size and converges on average within 6 to 7 iterations.

\section{Numerical Results and Discussion} 

We compare the results of our proposed \textbf{MaxNTtop} algorithm with those of Brute-force search by choosing the minimum transmit power allocation to any IoT node for firing its signal to it farthest neighboring node in the graph. The Brute-force algorithm will systematically enumerate all the possible solutions to the transmit power allocation problem in (\ref{eq17}) until all optimal solutions have been exhausted. Our proposed algorithm is also compared with the local minimal spanning tree (LMST) \cite{25} which is a topology control algorithm for communication graphs. We focus on a non-link-sharing scenario where each link is used only between a pair of nodes in a particular iteration.

\begin{figure}[t]
\begin{center}
 \includegraphics[width=0.5\linewidth]{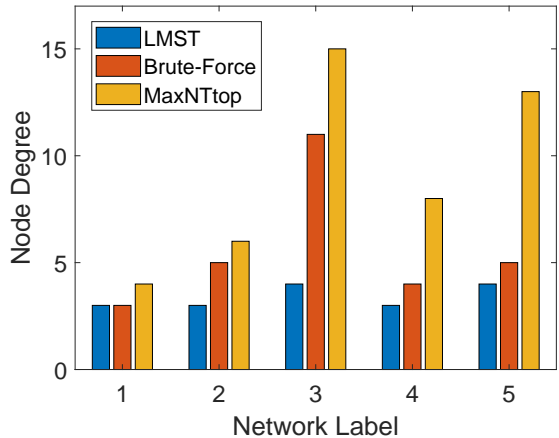}
\end{center}
\vspace*{-5mm}
\caption{Average comparative node degree achievable with different algorithms. A total of 5 random IoT networks are generated, each of which has 50 nodes distributed uniformly on an annulus surrounding a network gateway at the center with radius of 1 km.}
\label{FIG2}
\vspace*{-3mm}
\end{figure}

\begin{figure}[t]
\begin{center}
 \includegraphics[width=0.5\linewidth]{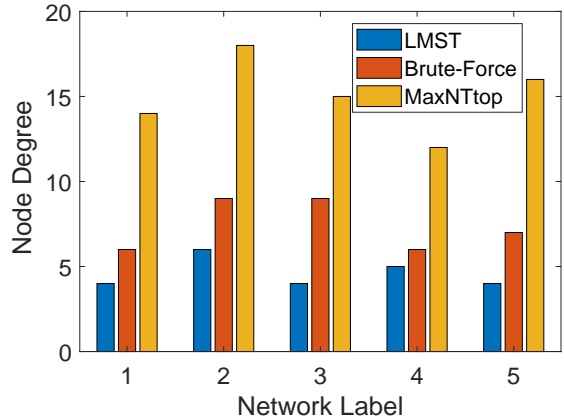}
\end{center}
\vspace*{-5mm}
\caption{Average comparative node degree achievable with different algorithms. A total of 5 random IoT networks are generated, each of which has 80 nodes randomly distributed over a $4\times 4$ km$^2$ rectangular area.}
\label{FIG3}
\vspace*{-3mm}
\end{figure}

The first set of results are generated for comparing the spatial usage capability offered by our algorithm in comparison to the LMST and Brute-force search. Spatial usage is defined as the network throughput of accommodating concurrent transmissions and is often measured in terms of node degree, i.e., the number of edges that can be connected to a node at the same time. So we are comparing average node degrees in Fig.~\ref{FIG2} for a total of 5 different IoT networks, each with 50 nodes that are distributed uniformly on an annulus surrounding a network gateway at the center with radius of 1 km. We considered average pathloss exponent of $\nu = 4$, $\rho_{R_{\text{min}}} = -63$ dBm, $\rho_{T_{\text{max}}} = 27$ dBm, $\beta = 2.5$ dB and a typical value of $\mathcal{N} = - 50$ dBm is considered for all simulation results. In Fig.~\ref{FIG3}, we increased the number of nodes per network to 80 nodes that are distributed randomly within a $4\times 4$ km$^2$ rectangular area without any central gateway. All the other simulation parameters are kept same as in Fig.~\ref{FIG2}.

\begin{figure}[t]
\begin{center}
 \includegraphics[width=0.5\linewidth]{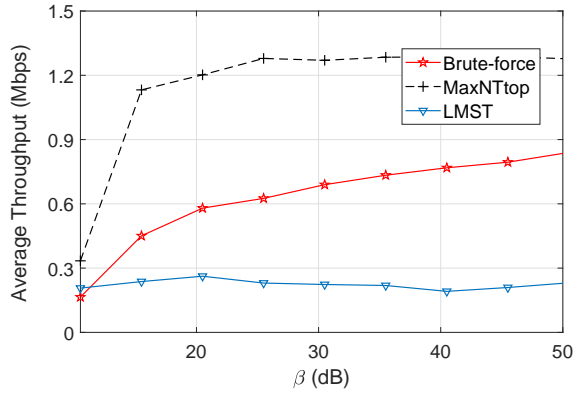}
\end{center}
\vspace*{-5mm}
\caption{Comparative achievable average throughput over different values of lower SNR threshold for an IoT network consisting of 50 nodes distributed uniformly on an annulus surrounding a network gateway at the center with radius of 1 km.}
\label{FIG4}
\vspace*{-3mm}
\end{figure}

\begin{figure}[t]
\begin{center}
 \includegraphics[width=0.5\linewidth]{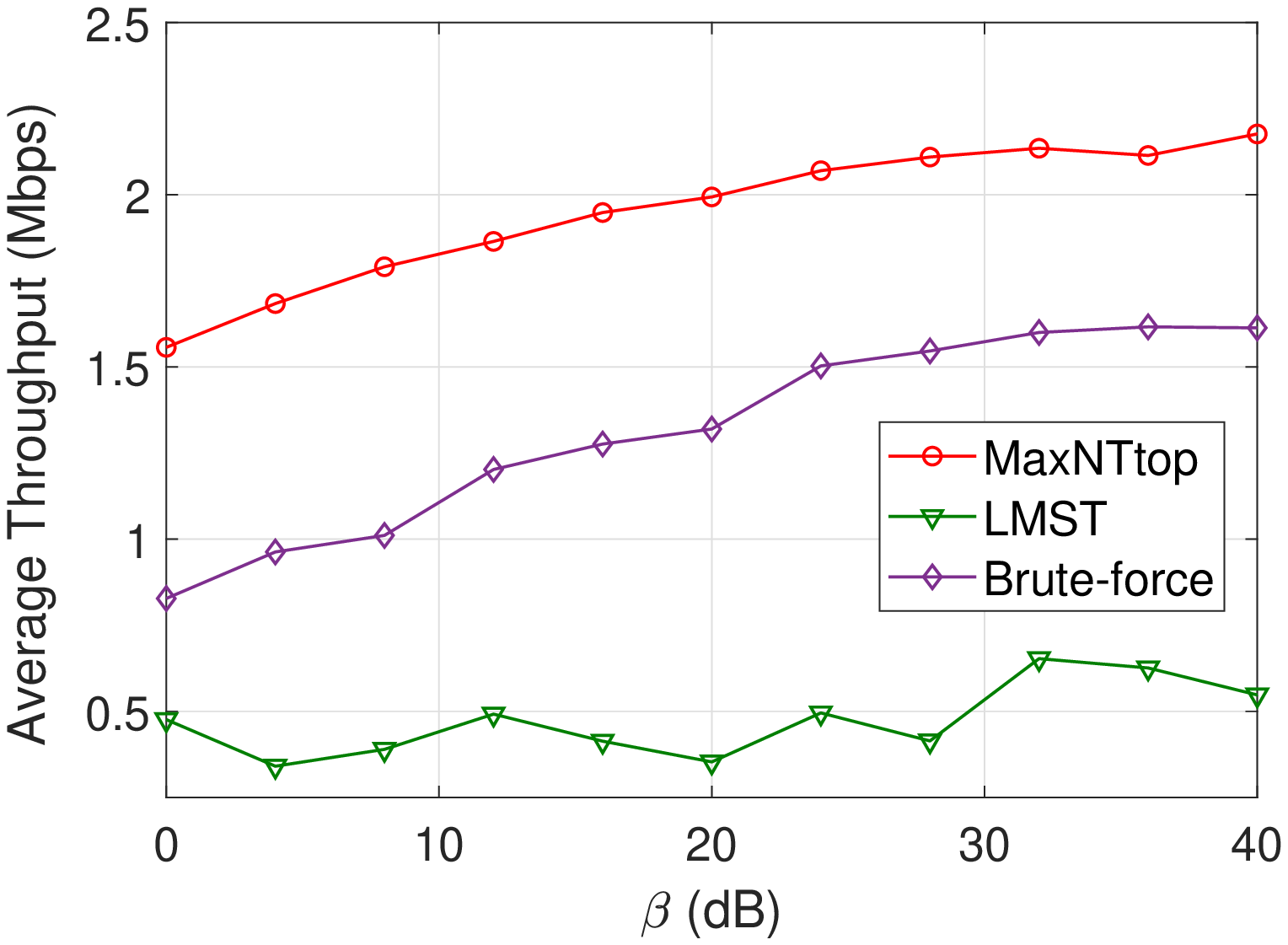}
\end{center}
\vspace*{-5mm}
\caption{Comparative achievable average throughput over different values of lower SNR threshold for an IoT network consisting of 80 nodes randomly distributed over a $4\times 4$ km$^2$ rectangular area.}
\label{FIG5}
\vspace*{-3mm}
\end{figure}
Both in Fig.~\ref{FIG2} and Fig.~\ref{FIG3}, LMST reports minimum average node degrees while Brute-force enables use of smaller number of nodes than the Brute-force search-based algorithm to cover a larger area in terms of higher possibility of connecting with more number of nodes. \textbf{MaxNTtop} offers best node degree both in the circular and rectangular spatial distribution. Higher spatial usage is observed when the number of nodes is increased in Fig.~\ref{FIG3}. This is due to the possibility of aligning patches within the entire network; \textbf{MaxNTtop} offers rotation and reflection of the patch to suitably align them in a way that patches sharing more than two common nodes can be sewed together. As a result a larger spatial area can be covered with a smaller number of IoT nodes. 

The second set of results in Fig.~\ref{FIG4} and Fig.~\ref{FIG5} are generated by comparing the average network throughput achievable by \textbf{MaxNTtop} with respect to LMST and Brute-force search. We use the same set of parameters used in Fig.~\ref{FIG2} and Fig.~\ref{FIG3} for Fig.~\ref{FIG4} and Fig.~\ref{FIG5} respectively. In this result set, we stitch patches with common nodes, one-by-one until all existing nodes within an IoT network are connected. Each new patch is sewed on the condition that it does not result in end-to-end violation of the minimum SNR constraint ($\beta$). From, Fig.~\ref{FIG4} and Fig.~\ref{FIG5}, we can see that \textbf{MaxNTtop} definitely improves network throughput. Now here we compare the network throughput against the SNR threshold and as the threshold increases, network throughput improves as all the links are experiencing higher average SNR. However, this improvement is capped and saturates when $\beta$ is increased over 25 dB. This limitation is observed as throughput is dependent on the link bandwidth. Also, intuitively, it gets more difficult to obtain connectivity with a higher SNR threshold to abide to. Therefore, even though the SNR constraint is allowed to increase, the average throughput cannot increase further as the link bandwidth is constant.

\section{Conclusions}

In this paper, we formulate optimized topology extraction and control for IoT networks, with the objective of maximizing network throughput. We model an IoT network as a global graph where a cluster of IoT nodes within a one-hop communication link is considered a patch or a sub-graph. The patches are aligned with each other when two patches share more than two common IoT nodes. The patches are aligned through eigenvector synchronization combining rotation, reflection and translation. After the patches are aligned, they are sewed together so that the topology created by sewing is capable of maximizing the coverage or spatial usage. Finally, the topology extracted in this way is optimized to offer maximum average network throughput under the physical signal-to-noise ratio (SNR) constraint. We demonstrated that our proposed algorithm \textbf{MaxNTtop} combining graph embedding and topology extraction offers improved network throughput and spatial usage, in comparison to Brute-force search and state-of-the-art LMST algorithms. Currently, \textbf{MaxNTtop} is a centralized approach. In future, it will be realistic to investigate how the topology extraction approach can be led in a distributed way such that each IoT node is intelligent enough to make their own decision on the neighbor to connect to, how much power to fire with or which direction to consider for information communications without relying on the central control gateway or any other node in the network.

\bibliographystyle{IEEEtran}

\end{document}